\documentclass[twocolumn,english,aps,letterpaper,prl,final]{revtex4}
\usepackage{times}
\usepackage[T1]{fontenc}
\usepackage[latin1]{inputenc}
\usepackage{geometry}
\usepackage{amsmath}
\usepackage{color}
\usepackage{graphicx}
\usepackage{amssymb}

\makeatletter
\usepackage{babel}
\makeatother
\begin{document}

\title{Two-body transients in coupled atomic-molecular BECs}

\author{\textcolor{black}{Pascal Naidon$\,^{1}$, Eite Tiesinga$\,^{1,2}$
and Paul S. Julienne$\,^{1,2}$}}

\affiliation{\textcolor{black}{$\,^{1}$Atomic Physics Division and $\,^{2}$Joint
Quantum Institute}\\
 \textcolor{black}{National Institute of Standards and Technology
and University of Maryland, 100 Bureau Drive Stop 8423, Gaithersburg,
Maryland 20899-8423, USA}}

\begin{abstract}
\textcolor{black}{We discuss the dynamics of an atomic Bose-Einstein
condensate when pairs of atoms are converted into molecules by single-color
photoassociation. Three main regimes are found and} it is shown that
they can be understood on the basis of time-dependent two-body theory.
In particular, the so-called rogue dissociation regime {[}Phys. Rev.
Lett., \textbf{88}, 090403 (2002)], which has a density-dependent
limit on the photoassociation rate, is identified with a transient
regime of the two-atom dynamics exhibiting universal properties. Finally,
we illustrate how these regimes could be explored by photoassociating
condensates of alkaline-earth atoms.\\
\\
\\

\end{abstract}
\maketitle
The conversion of atom pairs into molecules, using either Feshbach
resonances~\cite{kohler:1311} or photoassociation~\cite{jones:483},
serves as a tool to probe the many-body properties of ultracold gases~\cite{meiser:033621}.
In particular, photoassociation, the process of associating atoms
with a resonant laser light, was recently used to observe pair correlation
in a one-dimensional Bose gas~\cite{kinoshita:190406} an\textcolor{black}{d
the crossover between Bose-Einstein condensate (BEC) and Bardeen-Cooper-Schrieffer
(BCS) superfluid~\cite{partridge:020404}. Conversely, it can be
used to reach new regimes. Many-body theories have suggested the coherent
conversion of an atomic BEC into one of m}olecules~\cite{heinzen:5029,kostrun:063616},
macroscopic superposition~\cite{dannenberg:053601}, and production
of correlated atom pairs at high laser intensity~\cite{javanainen2002,gasenzer2004,naidon2005}.
Several experiments have made the first steps in these directions~\cite{mckenzie2002,prodan:080402,theis:123001,winkler2005},
but have been limited by inherent losses or insufficient laser power.
Reference~\cite{naidon2005} identified three regimes of photoassociation
as a function of loss and laser intensity. The intriguing density
dependence of the regime boundaries suggested that they are associated
to many-body effects.

In this letter, we first apply time-dependent two-body theory to photoassociation
with a single continuous laser and distinguish three transient regimes.
We then show how these transients explain the previously identified
regimes in the many-body theories. 

For two atoms of mass $M$ interacting with a resonant laser, photoassociation
is described by two equations coupling a scattering and a molecular
channel \cite{bohn1999},\begin{eqnarray*}
i\hbar\dot{\phi}(\vec{r})\!\!\! & = & \!\!\!\Big(-\frac{\hbar²}{M}\nabla²+U(r)\Big)\phi(\vec{r})+W(\vec{r})\phi_{m}(\vec{r})\\
i\hbar\dot{\phi_{m}}(\vec{r})\!\!\! & = & \!\!\!\Big(-\frac{\hbar²}{M}\nabla²+U_{m}(r)-i\frac{\gamma}{2}\Big)\phi_{m}(\vec{r})+W(\vec{r})\phi(\vec{r}),\end{eqnarray*}
where $\vec{r}$ is the relative separation of the two atoms, $\phi(\vec{r})$
and $\phi_{m}(\vec{r})$ are the components of the relative motion
wave function for the scattering and molecular channels, $U$ and
$U_{m}$ are the interaction potentials in each channel, $\gamma$
is the spontaneous emission rate from the molecular channel (we assume
that decayed molecules are lost from the system), and $W$ couples
the two channels. $W$ is proportional to the square root of the laser
intensity. We expand $\vert\phi\rangle$ and $\vert\phi_{m}\rangle$
in the bases of eigenstates of $-\frac{\hbar²}{M}\nabla²+U$ and $-\frac{\hbar²}{M}\nabla²+U_{m}$,
respectively, and assume that only the scattering eigenstates $\vert\varphi_{\vec{k}}\rangle$
(indexed by wave vector $\vec{k}$) are relevant in the scattering
channel, and that a single bound eigenstate $\vert\varphi_{m}\rangle$
is resonant in the molecular channel. Choosing $\langle\varphi_{m}\vert\varphi_{m}\rangle=1$
and $\langle\varphi_{\vec{k}}\vert\varphi_{\vec{p}}\rangle=(2\pi)^{3}\delta^{3}(\vec{k}-\vec{p})$,
one obtains\begin{eqnarray}
i\hbar\dot{C}_{\vec{k}}(t)\!\!\! & = & \!\!\! E_{k}C_{\vec{k}}(t)+w_{\vec{k}}C_{m}(t)\label{eq:TwoBodyExpanded1}\\
i\hbar\dot{C}_{m}(t)\!\!\! & = & \!\!\!(\Delta-i\gamma/2)C_{m}(t)+\int\!\!\!\frac{d^{3}\vec{k}}{(2\pi)^{3}}w_{\vec{k}}C_{\vec{k}}(t),\label{eq:TwoBodyExpanded2}\end{eqnarray}
where $C_{\vec{k}}$ and $C_{m}$ are the amplitudes in states $\vert\varphi_{\vec{k}}\rangle$
and $\vert\varphi_{m}\rangle$, $E_{k}=\hbar^{2}k^{2}/M$, $\Delta$
is the resonant bound state energy with respect to the scattering
threshold (which can be adjusted by tuning the laser frequency), and
$w_{\vec{k}}=\langle\varphi_{m}\vert W\vert\varphi_{\vec{k}}\rangle$
are the coupling matrix elements. According to Wigner's threshold
laws, $w_{\vec{k}}$ goes to a constant $w$ for low $k\ll1/\vert r_{c}\vert$,
where $r_{c}$ is the largest of the extent of the molecule, the van
der Waals length \cite{jones:483}, or the scattering length $a$
associated with $U$.

In ultracold gases, atoms collide at nearly zero energy. For the stationary
solution at zero energy, $C_{\vec{k}}$ goes to $-4\pi A/k^{2}$ for
low $k$, where $A$ is the optically-induced complex scattering length
\cite{fedichev1996,bohn1999} given by $4\pi\hbar^{2}A=-M\vert w\vert^{2}/(\Delta-\Delta^{\prime}-i\gamma/2)$
and the light shift $\Delta^{\prime}=\int\frac{d^{3}k}{(2\pi)^{3}}\frac{\vert w_{k}\vert^{2}}{E_{k}}$.
The imaginary part of $A$ is related to the loss rate coefficient\begin{equation}
K=-\frac{8\pi\hbar}{M}\textrm{Im}A=\frac{2}{\hbar}\textrm{Im}\frac{\vert w\vert^{2}}{\Delta-\Delta^{\prime}-i\gamma/2},\label{eq:StationaryRate}\end{equation}
which corresponds to the number of atoms lost per unit of time and
volume due to photoassociation to the excited state and subsequent
decay by spontaneous emission. From kinetic theory the density $\rho$
of remaining atoms in a thermal gas is expected to follow the rate
equation\begin{equation}
\dot{\rho}=-K\rho^{2}.\label{eq:RateEquation}\end{equation}

We now take into account the fact that the laser is turned on at $t=0$,
creating a strong perturbation to the two-body system. As a result,
transient regimes appear before the stationary solution is reached.
Initially, the two atoms are in the scattering channel with nearly
zero collision energy, \emph{i.e.} $C_{\vec{k}}=(2\pi)^{3}\delta^{3}(\vec{k})$
and $C_{m}=0$ at $t=0$. For $t>0$ we choose to decompose $C_{\vec{k}}(t)$
as follows

\begin{equation}
C_{\vec{k}}(t)=(2\pi)^{3}\delta^{3}(\vec{k})+C_{\vec{k}}^{ad}(t)+C_{\vec{k}}^{dyn}(t),\label{eq:Decomposition}\end{equation}
where $C_{\vec{k}}^{ad}(t)=-\frac{w_{\vec{k}}}{E_{k}}C_{m}(t)$ is
the adiabatic response to the turn-on of the laser (obtained by setting
$\dot{C}_{\vec{k}}=0$ in Eq.~(\ref{eq:TwoBodyExpanded1})), and
$C_{\vec{k}}^{dyn}(t)$ is the dynamic response. Solving for $C_{\vec{k}}^{dyn}$
from Eqs.~(\ref{eq:TwoBodyExpanded1}) and (\ref{eq:Decomposition})
and inserting it into Eq.~(\ref{eq:TwoBodyExpanded2}), we obtain\begin{eqnarray*}
i\hbar\dot{C}_{m}(t) & = & \Big(\Delta-\Delta^{\prime}-i\frac{\gamma}{2}\Big)C_{m}(t)\\
 &  & +w\Bigg[1+w{\displaystyle \frac{1-i}{2\hbar}\left(\frac{M}{2\pi\hbar}\right)^{3/2}}\int_{0}^{t}\frac{\dot{C}_{m}(\tau)}{\sqrt{t-\tau}}d\tau\Bigg],\end{eqnarray*}
where we have set $w_{\vec{k}}=w$, which is valid for $t\gg t_{c}$,
$t_{c}=Mr_{c}^{2}/\hbar$ being in most cases a very short time scale
on the order of 10~ns. We then assume that $\dot{C}_{m}(t)$ is localized
at short times, which leads to the Ansatz $\int_{0}^{t}d\tau\dot{C}_{m}(\tau)/\sqrt{t-\tau}=\alpha C_{m}(t)/\sqrt{t},$
where $\alpha$ is to be determined, and we approximate $i\hbar\dot{C}_{m}(t)$
by the short-time expression $i\hbar C_{m}(t)/t$. This leads to\begin{equation}
C_{m}(t)=\frac{-w}{\Big(\Delta-\Delta^{\prime}-i\frac{\gamma}{2}\Big)+S(t)-\frac{i\hbar}{t}},\label{eq:Cm}\end{equation}
where $S(t)=\frac{1-i}{2}\frac{\hbar}{\sqrt{t_{w}t}}$ and $\frac{i\hbar}{t}$
are time-dependent shifts and broadenings of the molecular level,
and $t_{w}=\frac{1}{\alpha^{2}}\left(\frac{h}{M}\right)^{3}\left(\frac{\hbar}{w}\right)^{4}$.
In the limit of large $S(t)$, one has $C_{m}(t)=-\frac{2w}{\hbar(1-i)}\sqrt{t_{w}t}$
which justifies that $\dot{C}_{m}(t)\propto t^{-1/2}$ is localized
at short times and yields $\alpha=\pi/2$. For small $S(t)$, our
Ansatz may not hold but this has little consequence precisely because
$S(t)$ is small.

We can now calculate an instantaneous rate coefficient, based on the
time variation of the population $\int_{k\approx0}\frac{d^{3}\vec{k}}{(2\pi)^{3}}\vert C_{\vec{k}}\vert^{2}$
in the initial state. This population is always inifinite, because
we started from a state which is not normalizable over an infinite
volume. However, from Eq.~(\ref{eq:TwoBodyExpanded1}), its time
derivative has a finite value, which we identify as minus the instantaneous
rate coefficient, $-K(t)$. Using Eq.~(\ref{eq:Decomposition}),
we can simplify it to \begin{equation}
K(t)=-\frac{2}{\hbar}\mbox{Im}\left(w^{*}C_{m}(t)\right).\label{eq:TimeDependentK}\end{equation}

\begin{figure}
\includegraphics[clip]{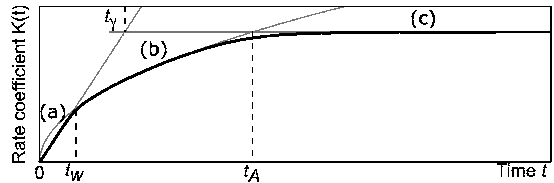}

\caption{\label{fig:evolution}Schematic evolution of the rate coefficient
(thick black curve) showing the three regimes~(a) ~(b), and~(c)
at sufficiently high laser intensity ($t_{w}\ll t_{A}$). The grey
curves correspond to the limiting expressions (\ref{eq:DynamicRate1}-\ref{eq:DynamicRate3}).
In particular, the parabolic curve associated with Eq.~(\ref{eq:DynamicRate2})
is a universal upper limit to the rate coefficient.}
\end{figure}

Depending on the relative strength of the terms in the denominator
of Eq.~(\ref{eq:Cm}), this coefficient goes through three subsequent
regimes illustrated in Fig.~\ref{fig:evolution}: linear (a), square
root (b), and constant with time (c). %
Namely, \begin{eqnarray}
\!\!\!\!\!\!\!\!\!(a)\quad K(t) & = & \frac{2}{\hbar}\frac{\vert w\vert^{2}}{\hbar}t,\qquad\qquad\mbox{for }t_{c}\ll t\ll t_{w}\label{eq:DynamicRate1}\\
\!\!\!\!\!\!\!\!\!(b)\quad K(t) & = & 4/\pi\left(h/M\right)^{3/2}\sqrt{t},\quad\mbox{for }t_{w}\ll t\ll t_{A}\label{eq:DynamicRate2}\\
\!\!\!\!\!\!\!\!\!(c)\quad K(t) & = & \frac{2}{\hbar}\textrm{Im}\frac{\vert w\vert^{2}}{\Delta-\Delta^{\prime}-i\gamma/2},\quad\mbox{for }t_{A}\ll t.\label{eq:DynamicRate3}\end{eqnarray}
where $t_{A}=\alpha^{2}\frac{2M}{h}\vert A\vert^{2}$. Note that $t_{w}\ll t_{A}$
only for high laser intensity. If $t_{w}\gtrsim t_{A}$, regime (a)
occurs for $t\ll t_{\gamma}$ and regime (c) occurs for $t\gg t_{\gamma}$,
where $t_{\gamma}=\hbar/\sqrt{(\Delta-\Delta^{\prime})^{2}+(\gamma/2)^{2}}.$

Photoassociation converts the initial state into the molecular state
and the loss rate coefficient grows linearly in regime~(a). In regime~(b)
the laser also drives molecules back to the atom-pair continuum. This
broadens the resonance and the loss rate coefficient still increases
but more slowly. Finally, in regime~(c) the molecules start decaying
spontaneously, and the rate coefficient reaches its steady-state maximal
value, which is the rate coefficient Eq.~(\ref{eq:StationaryRate})
obtained from the stationary solution.

The atomic density is expected to follow Eq.~(\ref{eq:RateEquation})
with the time-dependent rate coefficient~(\ref{eq:TimeDependentK}).
Whether the transient regimes (a) and (b) are observable depends on
whether an appreciable fraction of the density is photoassociated
over these time scales. Let us define the depletion time $t_{\rho}$
for which an appreciable fraction is depleted, \emph{i.e.} $K(t_{\rho})\rho_{0}t_{\rho}\approx1$,
where $\rho_{0}$ is the initial density. Three cases are possible.
If $t_{A},\, t_{\gamma}\ll t_{\rho}$, then only the final constant
rate coefficient is relevant. If $t_{w}\ll t_{\rho}\ll t_{A}$, then
the first regime to lead to observable losses is regime (b). If $t_{\rho}\ll t_{w},\, t_{\gamma}$,
then the only relevant regime is (a).

Interestingly, when regime (b) dominates, the molecular amplitude
vanishes and the system has a universal behavior. The loss rate (\ref{eq:DynamicRate2})
and the atom pair distribution\[
C_{\vec{k}}(t)=(2\pi)^{3}\delta^{3}(\vec{k})+\frac{4}{k^{2}}\sqrt{\frac{2iht}{M}}-\frac{4\pi i}{k^{3}}\mbox{Erf}\Big(k\sqrt{\frac{\hbar t}{iM}}\Big)e^{-ik^{2}\frac{\hbar t}{M}}\]
do not depend on the microscopic details of the transition, but just
on the mass of the species. Here, Erf denotes the error function.
The condition $t_{w}\ll t_{\rho}\ll t_{A}$, needed for the observation
of this regime, is equivalent to\begin{equation}
\rho_{0}^{1/3}\vert A\vert\gg(2\pi)^{-2/3}\quad\mbox{ and }\quad\rho_{0}^{1/3}B\ll(2\pi)^{-2/3}\label{eq:condition12}\end{equation}
where the length $B$ is $8\left(\frac{\hbar^{2}}{Mw}\right)^{2}$.
Finite temperature adds the condition that $K(t_{\rho})$ be smaller
than the unitarity limit $\frac{h}{M}\lambda$, where $\lambda$ is
the de Broglie wavelength. It fortuitously coincides with the condensation
condition for bosons $\rho\lambda^{3}\gtrsim1$ \cite{javanainen2002,mckenzie2002}.

\begin{figure}
\includegraphics{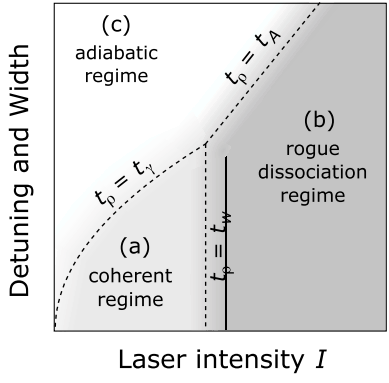}

\caption{\label{Regimes}Regimes of photoassociation in a BEC as a function
of the {}``detuning and width'' $\sqrt{(\Delta-\Delta^{\prime})^{2}+(\gamma/2)^{2}}$
and the laser intensity $I$, for a fixed density. The boundaries
are indicated by dashed lines and equalities of time scales defined
in the text.}
\end{figure}

We now turn to a many-body description. Photoassociation in a uniform
BEC can be described (up to first order in a cumulant expansion \cite{kohler2002})
by the three equations \cite{gasenzer2004,naidon2005}\begin{eqnarray*}
i\hbar\dot{\Psi}\!\!\! & = & \!\!\!\Psi^{*}\int\!\Big(U(r)\Phi(\vec{r})+W(r,t)\Phi_{m}(\vec{r})\Big)d³r\\
i\hbar\dot{\Phi}(\vec{r})\!\!\! & = & \!\!\!\Big(\!\!-\!\!\frac{\hbar²}{M}\nabla²\!+\! U(r)\!\Big)\Phi(\vec{r})+W(\vec{r})\Phi_{m}(\vec{r})+2i\hbar\Psi\dot{\Psi}\\
i\hbar\dot{\Phi}_{m}(\vec{r})\!\!\! & = & \!\!\!\Big(\!\!-\!\!\frac{\hbar²}{M}\nabla²\!+\! U_{m}(r)\!-i\gamma/2\Big)\Phi_{m}(\vec{r})+W(\vec{r})\Phi(\vec{r})\end{eqnarray*}
where $\Psi$ is the condensate wavefunction (here a complex number),
and $\Phi(\vec{r})$ and $\Phi_{m}(\vec{r})$ are the pair wavefunctions
in the scattering and molecular chann\textcolor{black}{els. Higher-order
cumulants (such as the normal density of noncondensate atoms \cite{holland2001})
contribute significantly only for $t\gtrsim t_{\rho}$, and do not
change the dynamics up to 10 $\mu$s in the examples to follow. Inelastic
collisions \cite{yurovsky2003}, not included here, do not affect
the atomic BEC during this time frame for typical rate coefficients
$\sim10^{-10}\mbox{ cm}^{3}\mbox{s}^{-1}$.}

\textcolor{black}{As in the two-body case, we can write}\begin{eqnarray*}
\Phi(\vec{r}) & = & \Psi^{2}+\int\!\!\frac{d^{3}\vec{k}}{(2\pi)^{3}}(C_{\vec{k}}^{ad}+C_{\vec{k}}^{dyn})\varphi_{\vec{k}}(\vec{r})\\
\Phi_{m}(\vec{r}) & = & \Psi_{m}\varphi_{m}(\vec{r}),\end{eqnarray*}
where the adiabatic part $C_{\vec{k}}^{ad}=-\frac{1}{E_{k}}(w_{\vec{k}}\Psi_{m}+g_{\vec{k}}\Psi^{2})$
and $g_{\vec{k}}=\langle0\vert U\vert\varphi_{\vec{k}}\rangle$, $\vert0\rangle$
being the zero-momentum plane wave. Elimination of the adiabatic part
is now crucial, because it introduces the light shift $\Delta^{\prime}$
but also the coupling constants $w_{\vec{k}}$ and $g_{\vec{k}}$
without resorting to \textcolor{black}{contact interactions \cite{naidon2005}.
We find\begin{eqnarray}
\!\!\!\!\!\!\!\!\! i\hbar\dot{\Psi}\!\! & = & \Psi^{*}\Big(g_{\vec{0}}\Psi²+\int\!\!\frac{d³\vec{k}}{(2\pi)³}g_{\vec{k}}C_{\vec{k}}^{dyn}+w_{\vec{0}}\Psi_{m}\Big)\label{eq:manybody1}\\
\!\!\!\!\!\!\!\!\! i\hbar\dot{\Psi}_{m}\!\! & = & \!\!\Big(\!\Delta\!-\!\Delta^{\!\prime}\!-\! i\frac{\gamma}{2}\Big)\!\Psi_{m}\!+\!\left(\! w_{\vec{0}}\Psi^{2}\!\!+\!\!\!\int\!\!\!\frac{d³\vec{k}}{(2\pi)³}w_{\vec{k}}C_{\vec{k}}^{dyn}\!\right)\label{eq:manybody2}\\
\!\!\!\!\!\!\!\!\! i\hbar\dot{C}_{\vec{k}}^{dyn}\!\! & = & E_{k}C_{\vec{k}}^{dyn}-i\hbar\dot{C}_{\vec{k}}^{ad}.\label{eq:manybody3}\end{eqnarray}
These equations are similar to those of Ref.~\cite{javanainen2002},
but contain no ultraviol}et divergence. Thus, we can thus safely set
$w_{\vec{k}}=w$ and $g_{\vec{k}}=g=4\pi\hbar^{2}a/M$ without any
renormalization.

\textcolor{black}{When the dynamical part $C_{k}^{dyn}$ is negligible
we obtain the familiar set of coupled Gross-Pitaevski\u{\i} equations
introduced in Ref.~\cite{timmermans1999}. If we assume that $\Psi=\sqrt{\rho_{0}}$
and $\Psi_{m}=0$ initially, these equations admit two limiting regimes
\cite{timmermans1999,naidon2005}. In the adiabatic regime, $w^{2}\rho_{0}\ll(\Delta-\Delta^{\prime})^{2}+(\gamma/2)^{2}$,
the molecular wave function can be adiabatically eliminated, and the
condensate density $\rho=\vert\Psi\vert^{2}$ then follows Eq.~(\ref{eq:RateEquation})
with the rate coefficient (\ref{eq:StationaryRate}) predicted by
the stationary two-body theory. On the other hand, in the coherent
regime, $w^{2}\rho_{0}\gg(\Delta-\Delta^{\prime})^{2}+(\gamma/2)^{2}$,
$\Psi$ and $\Psi_{m}$ exhibit coherent Rabi oscillations at a frequency
$\Omega=w\sqrt{\rho_{0}}/\hbar$.}

\begin{figure}
\includegraphics[clip,scale=0.54]{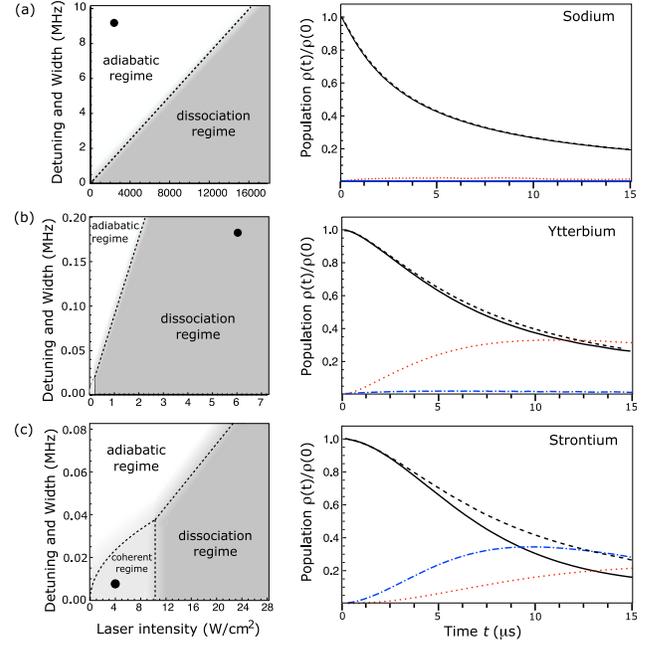}

\caption{\label{fig:ensemble}(color online) On-resonance photoassociation
($\Delta-\Delta^{\prime}=0$) of a condensate of sodium (a), ytterbium
(b) and strontium (c) for typical transitions. In each row, the right
panel shows the population evolutions, according to Eqs.~(\ref{eq:manybody1}-\ref{eq:manybody3}),
based on the parameters indicated by the black dot in the left panel,
which is a regime diagram similar to Fig.~\ref{Regimes}. Solid curve:
atomic condensate population; dot-dashed curve: molecular population;
dotted curve: correlated pair population. The short-dashed curve shows
the atomic condensate population following from Eq.~(\ref{eq:RateEquation})
with the rate (\ref{eq:EffectiveRate}). For sodium, we used the conditions
of Ref.~\cite{mckenzie2002} ($\vert\mbox{Im}A\vert/I=2.95\;\mbox{fm}/(\mbox{W}\mbox{cm}^{-2})$,
$\gamma/\hbar=18$ MHz). Typical intercombination transition parameters
were used for ytterbium \cite{tojo2006} ($\vert\mbox{Im}A\vert/I=2.12\;\mbox{nm}/(\mbox{W}\mbox{cm}^{-2})$
, $\gamma/\hbar=364$ kHz) and strontium \cite{zelevinsky2006} ($\vert\mbox{Im}A\vert/I=2.12\,\mbox{nm}/(\mbox{W}\mbox{cm}^{-2})$,
$\gamma/\hbar=15$ kHz). Thus, we have $(t_{w},\, t_{\gamma},\, t_{A})=(0.17,\,0.017,\,0.0036)\,\mu\mbox{s}$
for sodium, $(0.044,\,0.88,\,35)\,\mu\mbox{s}$ for ytterbium, and
$(116,\,21,\,7.8)\,\mu\mbox{s}$ for strontium. For all cases, the
initial density is $\rho_{0}=6\cdot10^{14}\mbox{ cm}^{-3}$, and $t_{\rho}\sim5\,\mu$s.}
\end{figure}

The case where the dynamical part $C_{\vec{k}}^{dyn}$ cannot be neglected
corresponds to the {}``rogue dissociation limit'' of Ref.~\cite{javanainen2002}.
Ref.~\cite{naidon2005} showed that this occurs when\begin{equation}
\rho_{0}^{1/3}\vert A\vert\gg\frac{1}{2}\left(\frac{\pi}{2}\right)^{1/3}\;\;\!\!\mbox{ and }\!\!\quad\rho_{0}^{1/3}L\gg\frac{1}{2}\!\!\left(\!\frac{\pi}{2}\!\right)^{1/3}\label{eq:rogue12}\end{equation}
for the adiabatic and coherent regimes, respectively, where $L=\frac{M}{4\pi\hbar^{2}}\frac{w}{\sqrt{2\rho_{0}}}$
was identified with a many-body length. Figure~\ref{Regimes} shows
the three regimes: adiabatic, coherent and rogue dissociation. Most
intriguing has been the dependence of the regime boundaries on the
density. In particular, increasing the density makes it easier to
reach rogue dissocation from the adiabatic regime, but more difficult
from the coherent regime, according to Eqs.~(\ref{eq:rogue12}). 

In light of the previous two-body analysis, we can now interpret the
adiabatic, coherent and rogue regimes in terms of the two-body regimes
(a), (b) and (c). Using the same approximations as in the two-body
theory, we can reduce the many-body equations (\ref{eq:manybody1}-\ref{eq:manybody3})
to a rate equation with a time-dependent coefficient\begin{equation}
K(t)=\frac{2}{\hbar}\textrm{Im}\frac{\vert w\vert^{2}+gS(t)}{\Delta-\Delta^{\prime}-i\frac{\gamma}{2}+S(t)-\frac{i\hbar}{t}}.\label{eq:EffectiveRate}\end{equation}
We emphasize that this rate coefficient does not depend on the density
and originates essentially from the time-dependent two-body coefficient
(\ref{eq:TimeDependentK}). It follows that the three regimes (a),
(b) and (c) also apply for (\ref{eq:EffectiveRate}). In fact, the
conditions for the observation of regime (a), (b) or (c) during the
depletion time $t_{\rho}$, expressed in terms of $t_{w}$, $t_{\gamma}$,
and $t_{A}$ are equivalent to the boundary conditions for respectively
the coherent, rogue dissociation, and adiabatic regime, as shown in
Fig.~\ref{Regimes}. Thus, the density dependence of these boundaries
originates simply from the usual rate equation of kinetic theory.
Note in particular that the rogue regime boundaries (\ref{eq:rogue12})
are equivalent to the conditions (\ref{eq:condition12}) for the observation
of the universal regime (b). This shows that the length $L$ has in
fact no special significance. 

The right panels of Fig.~\ref{fig:ensemble} compare the atomic condensate
evolution from Eqs.~(\ref{eq:manybody1}-\ref{eq:manybody3}) with
that following from the rate coefficient Eq.~(\ref{eq:EffectiveRate})
in the three regimes. It shows that the short-time evolution is always
consistent with two-body dynamics and kinetic theory, justifying the
approximations we have made. Only in the coherent regime, for $t\gtrsim t_{\rho}$,
is the condensate nature of the gas revealed due to collective effects
which cannot be described by a rate equation.

For \textcolor{black}{alkali-metal atoms experimentally studied in
\cite{mckenzie2002,winkler2005}, the molecular state has a short
lifetime on the order of 10 ns. As a result, these experiments have
been confined to the adiabatic regime. Figure~\ref{fig:ensemble}a
shows a typical case for sodium. To reach other regimes, very large
intensities are needed. On the other hand, photoassociation near narrow
intercombination lines leads to much longer-lived molecules. Figures~\ref{fig:ensemble}b
and \ref{fig:ensemble}c show on-resonance photoassociation in condensates
of ytterbium and strontium, for typical states below the intercombination
line. For moderate intensities, it appears possible to reach the universal
regime (b) of pair dissociation. This creates correlated pairs of
atoms from a condensate, analagous to correlated photons \cite{PhysRevLett.86.3180}.
For strontium, it should be possible to reach the coherent regime,
at least partially. Note, however, that inelastic collisions, while
not significantly affecting the atomic condensate, reduce the molecular
and noncondensate populations by about a half in Fig.~\ref{fig:ensemble}c.}

\textcolor{black}{In summary, we showed how different regimes of photoassociation
in a BEC originate from transient shifts a}nd broadenings in the two-atom
dynamics. These have simple analytical expressions, and lead to a
universal behavior in the rogue/pair dissociation regime.

We thank the Office of Naval Research for partial support. \bibliographystyle{apsrev}
\bibliography{bibliopaper}

\end{document}